\documentclass[twocolumn,prb]{revtex4}% Physical Review B
\usepackage{graphicx}% Include figure files
\usepackage{dcolumn}% Align table columns on decimal point
\usepackage{bm}% bold math

\begin{document}

\title{Cooling rate dependence of the antiferromagnetic domain structure of a single crystalline charge ordered manganite.}

\author{R. Mathieu and P. Nordblad}
\affiliation{Department of Materials Science, Uppsala University, Box 534, SE-751 21 Uppsala, Sweden}

\author{A. R. Raju and C. N. R. Rao}
\affiliation{Chemistry and Physics of Materials Unit, 
Jawaharlal Nehru Centre for Advanced Scientific Research,\\Jakkur P.O., Bangalore - 560 064, India}

\date{\today}

\begin{abstract}
The low temperature phase of single crystals of Nd$_{0.5}$Ca$_{0.5}$MnO$_3$ and Gd$_{0.5}$Ca$_{0.5}$MnO$_3$ manganites is investigated by squid magnetometry. Nd$_{0.5}$Ca$_{0.5}$MnO$_3$ undergoes a charge-ordering transition at $T_{CO}$=245K, and a long range CE-type antiferromagnetic state is established at $T_N$=145K. The dc-magnetization shows a cooling rate dependence below $T_N$, associated with a weak spontaneous moment. The associated excess magnetization is related to uncompensated spins in the CE-type antiferromagnetic structure, and to the presence in this state of fully orbital ordered regions separated by orbital domain walls. The observed cooling rate dependence is interpreted to be a consequence of the rearrangement of the orbital domain state induced by the large structural changes occurring upon cooling.

\end{abstract}

\maketitle

%\begin{multicols}{2}

\section{Introduction}
Hole-doped manganites R$_{1-x}$A$_x$MnO$_3$ (R=La,Nd,Pr,... and A=Sr,Ca,...), exhibit colossal magnetoresistive\cite{ramirez,rao} (CMR) properties, associated with the mixed manganese valence Mn$^{3+}$ ($t^{3}_{2g}e^{1}_g$)/Mn$^{4+}$($t^{3}_{2g}$) resulting from the substitution of trivalent R ions by $x$ divalent A ions. The $x$ $\sim$ 0.5 doping is particularly interesting, since the magnetic interaction is affected by i) the ordering of the Mn$^{3+}$ and Mn$^{4+}$ charges\cite{rao,kruka}, commonly referred as charge-ordering (CO), ii) the ordering of the $e_g$ electron orbitals on the Mn$^{3+}$ sites and iii) the coupling between orbital degree of freedom and the lattice. For example, Nd$_{0.5}$Ca$_{0.5}$MnO$_3$ (NCMO) undergoes a CO transition\cite{rao-oldncmo} at $T_{CO}$=245K with partial orbital ordering (OO) and magnetic correlations of short range. At lower temperatures, the OO increases and a long range CE-type\cite{good} antiferromagnetic (AFM) state is established at $T_N$=145K. An insulator-to-metal transition occurs around this temperature in intermediate ($\sim$5T or higher) magnetic fields\cite{rao-oldncmo}. The CE-type phase involves both FM and AFM interactions. It consists of ferromagnetic (FM) chains in the (a,b) plane, AFM coupled between each other within this plane, and along the $c$-axis. Neutron\cite{cox} and Brillouin\cite{rao-ncmo} scattering experiments detect additional FM correlations in the low temperature CE-type AFM phase. Also resistivity noise measurements reveal two-level fluctuations\cite{noise1} related to the phase separation scenario\cite{elbio}, including mixed-phase states of different magnetic and electrical properties. In this context, the observed magnetic field induced metal-insulator transition\cite{tokura1} reflects the percolative growth of conducting FM clusters embedded in an AFM matrix\cite{noise2}.\\
In the present article, we investigate the stability of the low temperature AFM phase of a Nd$_{0.5}$Ca$_{0.5}$MnO$_3$ single crystal using ac and dc magnetization measurements. The zero field cooled (ZFC) and field cooled (FC) magnetizations are recorded vs. temperature and time after specific cooling protocols. Similar measurements are performed on a Gd$_{0.5}$Ca$_{0.5}$MnO$_3$ (GCMO) single crystal for comparison. GCMO also shows charge-ordering at $T_{CO}$ = 260K, but no long range AFM, and it remains insulating at all temperatures, even in large magnetic fields\cite{rao-xcmo}. In the case of NCMO, the magnetization is cooling rate dependent below $T_N$, and a weak spontaneous moment appears in the AFM state. The corresponding excess magnetization appearing along the zig-zag chains of the CE-Type structure is related to the presence of domain walls in the (a,b) plane breaking the orbital coherency.

\section{Samples and experiments}
Single crystals of NCMO\cite{rao-ncmo} and GCMO were grown using a floating zone furnace (NEC, Japan). Temperature and time dependent zero field cooled and field cooled magnetization measurements were performed using a Quantum Design MPMS5 Superconducting QUantum Interference Device (SQUID) magnetometer. In the ZFC and FC case, the magnetization $M$ was collected on re-heating in a small magnetic field $H$=20 Oe after slow (3K/min) and fast (60K/min) cooling from room temperature down to 5K. $M$ vs. $H$ measurements were performed at $T$=35K after similar cooling protocols. Additional ac-susceptibility $\chi$ measurements ($f$=125Hz, $h$=20 Oe) were recorded using the same cooling protocol on a Lakeshore 7225 susceptometer for comparison. The volume susceptibility, $M/H$ in dc and $\chi$ in ac, is in the following plotted in dimensionless (SI) units. The correspondence between $M/H$ in (SI) units and $M$ in bohr magnetons per formula unit ($\mu_B$/f.u) is indicated in the text and figures.

\section{results and discussion}
\begin{figure}
\includegraphics[scale=0.40]{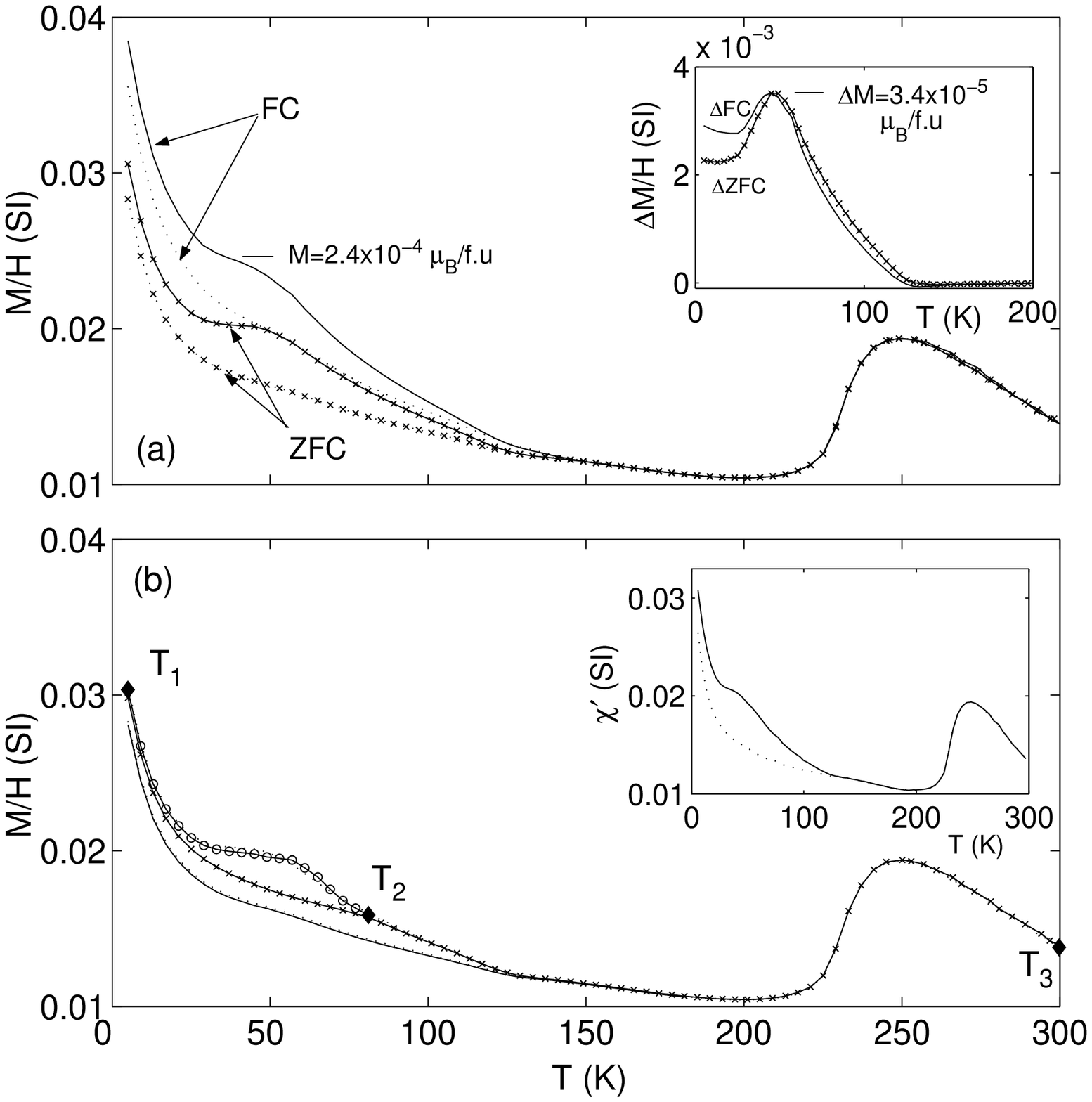}
\caption{Temperature dependence of the ZFC (markers) and FC (simple line) magnetizations of NCMO; $H$=20 Oe. (a) The magnetization is recorded along the c-axis on re-heating after fast (continuous line) and slow (dotted line) cooling down to 5K. The corresponding value of $M_{FC}$($T$=35K) in $\mu_B$/f.u is added for comparison. The insert shows the difference plots of the FC and ZFC magnetization curves (same symbols as in the main frame). The maximum value of $\Delta M$ in $\mu_B$/f.u is also indicated. (b) Idem adding a cooling and re-heating to a temperature $T_2$ below $T_N$ while recording $M_{ZFC}$($T$). $T_1$, $T_2$ and $T_3$ correspond to $T$=5, 80 and 300K respectively; see main text. The insert shows the cooling rate dependence of the ac-susceptibility of NCMO, recorded vs. temperature using a small ac-field, after similar fast and slow cooling. $h$=20 Oe, $f$=125Hz. The simple line corresponds to fast cooling, and the dotted line to slow cooling.}
\label{fig1}
\end{figure}

\begin{figure}
\includegraphics[scale=0.40]{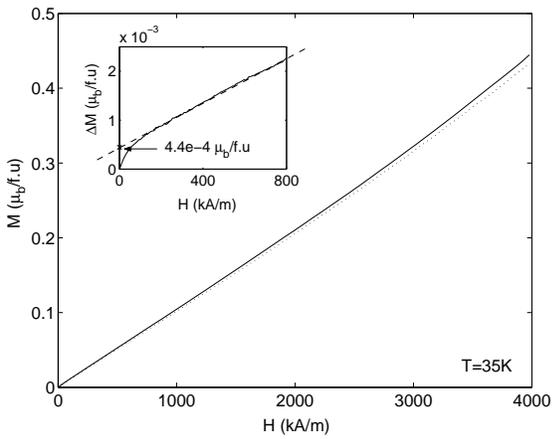}
\caption{$M$ vs. $H$ up to high magnetic fields recorded after fast (continuous line) and slow (dotted line) cooling of NCMO; $T$=35K. The insert shows the corresponding $\Delta M$ [=$M(fast$ $cooling)-M(slow$ $cooling)$] difference plot.}
\label{fig2}
\end{figure}

\begin{figure}
\includegraphics[scale=0.40]{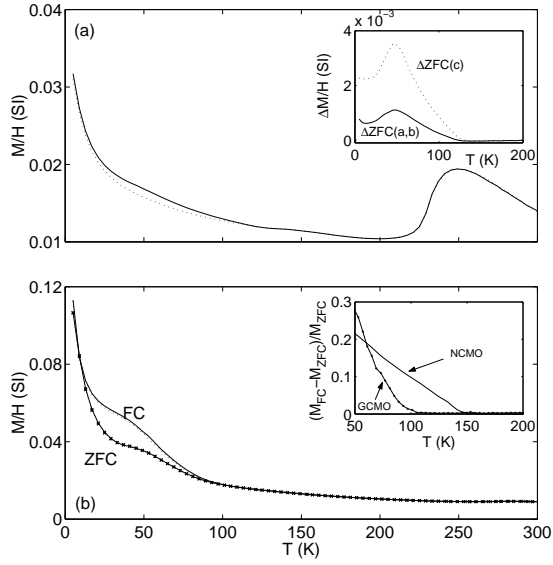}
\caption{(a) Temperature dependence of the ZFC magnetization for NCMO. The magnetization in the (a,b)-plane is recorded on re-heating after fast (continuous line) and slow (dotted line) cooling to 5K. The insert shows the corresponding difference plot. The curve obtained when measuring along the c direction (c.f. insert of Fig. 1(a)) is added. (b) Corresponding results for GCMO. $M$ is measured along the c-axis and both ZFC (markers) and FC (simple line) are measured with fast (continuous line) and slower (dotted line) cooling rates. The insert shows the temperature dependence of the irreversibility observed in the magnetization curves of NCMO and GCMO.}
\label{fig3}
\end{figure}

\begin{figure}
\includegraphics[scale=0.40]{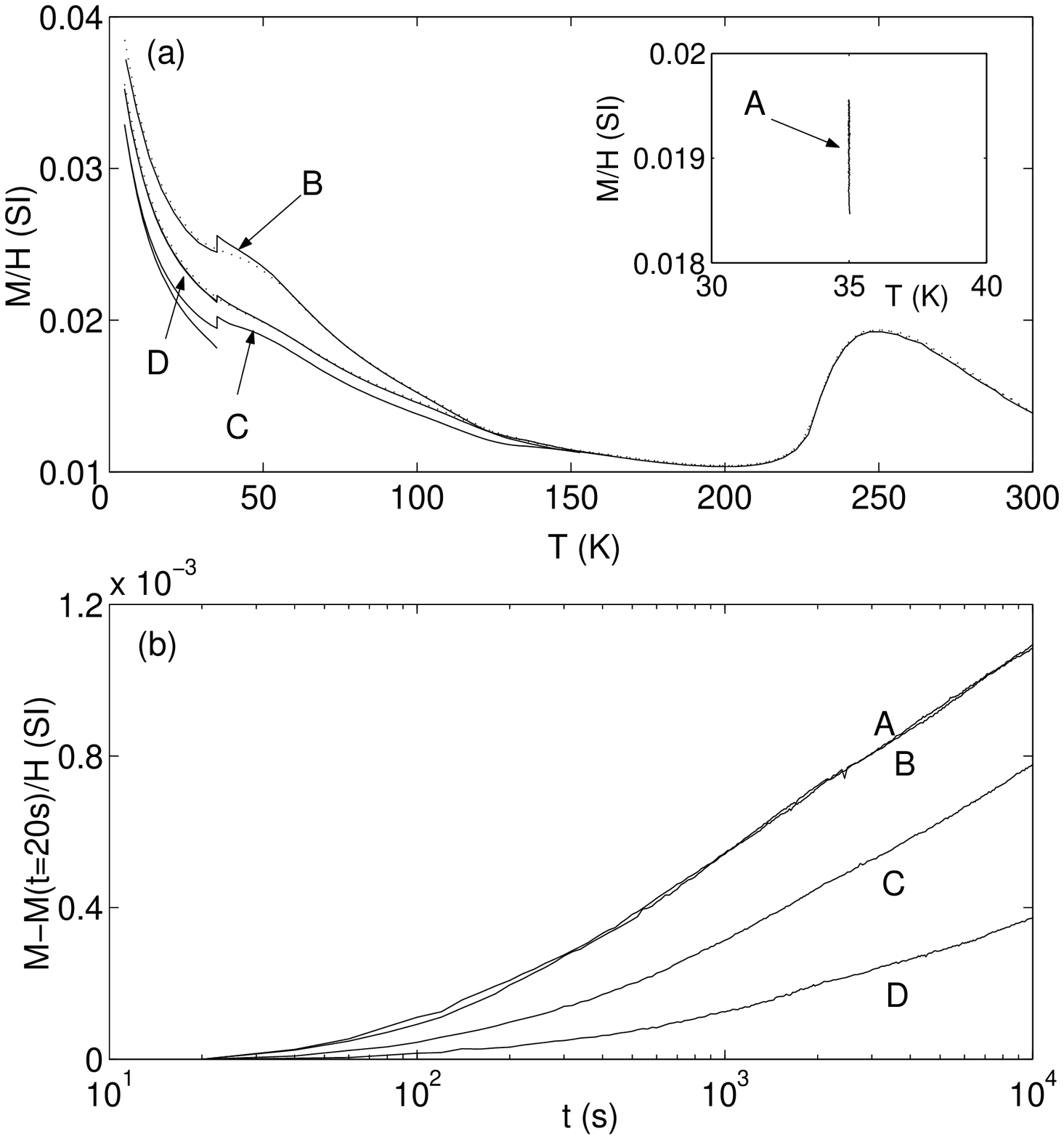}
\caption{Temperature dependence of the $M_{FC}$ for NCMO. Intermediate temperature stops are made during the cooling and heating at $T_{m}$=35K (see main text) (a) shows the results plotted vs. temperature, as well as the reference $M_{FC}$($T$) curves from Fig. 1(a) in dotted lines. Curve \textbf{A} is plotted separately for clarity; $M_{FC}$ was only recorded vs. time in this experiment. (b) shows the the relaxation of $M_{FC}$ during $t_m$ at $T_m$=35K.}
\label{fig4}
\end{figure}

Figure \ref{fig1} shows the cooling rate dependence of the ZFC and FC magnetization for NCMO. (a) shows the temperature dependence of the ZFC (markers) and FC (simple line) magnetization, measured on re-heating after fast (continuous lines) and slow (dotted) cooling. In the case of fast cooling or quench to low temperatures, an excess magnetization $\Delta M$ appears below $T_N$, both in the ZFC and FC curves. Difference plots of the FC and ZFC curves (same symbols as in the main frame) are added in the insert, showing that the excess magnetization $\Delta M$ [=$M(fast$ $cooling)-M(slow$ $cooling)$] appears slightly below $T_N$, around $T$=130K. This excess magnetization relaxes with time, as illustrated in Fig. \ref{fig1}(b) which shows the temperature dependence of the ZFC magnetization $M_{ZFC}$($T$): As in Fig. \ref{fig1}(a), $M_{ZFC}$ is recorded on reheating from the lowest temperature $T_1$=5K up to a temperature $T_2$=80K below $T_N$ after a fast cooling. The $M_{ZFC}$($T_1 \rightarrow T_2$) curve is marked using circles. The sample is cooled back (3K/min) to $T_1$ and $M$ is recorded on re-heating up to $T_3$=300K, well above $T_{N}$; the $M_{ZFC}$($T_1 \rightleftharpoons T_2 \rightarrow T_3$) curve is marked using crosses. The sample is cooled down (3K/min) to $T_1$, and $M$ is once again recorded up to $T_3$; the $M_{ZFC}$($T_1 \rightarrow T_3$) curve is plotted using a simple line. The $M_{ZFC}$($T$) curves obtained in Fig. \ref{fig1}(a) are added as dotted lines for comparison. As expected, both the first (from $T_1$ to $T_2$ after a fast cooling) and the last (from $T_1$ to $T_3$ after a slow cooling) measurements coincide with the earlier results shown in Fig. \ref{fig1}(a). The second measurement, recorded from $T_1$ to $T_3$ after a temperature cycle to $T_2$, yields instead a lower $\Delta M$, showing that the excess magnetization is relaxing with time. It should be stressed that only a small dc-magnetic field, acting essentially as a non perturbing probe of the system, is used to record the magnetization; i.e. the observed effects are not driven by the magnetic field employed in the experiments\cite{note}. To evidence this, we show in insert of Fig. \ref{fig1}(b) the temperature dependence of the in-phase component of the ac-susceptibility recorded under the same conditions as the ZFC and FC magnetization, after fast (simple line) and slow (dotted line) cooling; a similar cooling rate dependence is observed in the ac-susceptibility.\\
In our weak probing field, $\Delta M$ amounts to 2.9$\times$10$^{-5}$ $\mu_B$/f.u at $T$=35K (c.f.  Fig. \ref{fig1}(a)). $M$ vs. $H$ measurements up to higher fields (4000 kA/m) recorded after fast (continuous line) and slow (dotted line) cooling to $T$=35K are shown in Figure \ref{fig2}. In both cases, a closely linear field dependence of the magnetization is observed, reflecting the AFM order. The difference plot of the two curves (see insert) reveals a weak spontaneous moment of 4.4$\times$10$^{-4}$ $\mu_B$/f.u., superposed on a small excess susceptibility. The small moment reflects the presence of uncompensated spins in the AFM state. The origin of these local defects of the magnetic structure will be discussed below.\\
In the above described measurements, $M$ was recorded using a magnetic field directed along the $c$-axis of NCMO. Figure \ref{fig3}(a) shows the corresponding results on the ZFC magnetization measured in the (a,b) plane. A cooling rate dependence is again observed, but the excess magnetization has a smaller magnitude, as seen in the main frame and also in the insert where the difference plots of the ZFC curves for the two different orientations of $H$ are shown. Most of the excess magnetization thus lies along the $c$ direction. The single crystals show no sign of twinning, but of course some magnetic anisotropy in the (a,b) plane would affect the magnitude of $\Delta M$.\\
In the case of GCMO, a cooling rate dependence is not observed, and the magnetizations curves recorded along $c$ after slow and fast cooling virtually coincide, as shown in Fig. \ref{fig3}(b). On the other hand, the ZFC and FC curves deviate from each other below $T\sim$100K, indicating irreversibility below this temperature, and thus the development of magnetic correlation. The insert of Fig. \ref{fig3}(b) shows the normalized difference between the ZFC and FC curves for both GCMO and NCMO, which is a measure of the irreversibility. In the case of NCMO, the irreversibility arises sharply at $T_N$, while it appears more gradually in GCMO. The magnetic correlation developing below $T\sim$ 100K in GCMO seems to remain of short range, as in the very similar Y$_{0.5}$Ca$_{0.5}$MnO$_3$\cite{raoycmo} manganite. The excess magnetization, and the observed cooling rate dependence of the magnetization at low temperatures in NCMO are instead related to the establishment of the long range AFM state, possibly via the large increase in orthorhombic distortion\cite{rao-xcmo} and $c$-axis contraction\cite{rao-oldncmo} occurring between $T_{CO}$ and $T_N$ upon cooling. The spontaneous moment could then be related to the presence of defects in the low temperature antiferromagnetic arrangement, or an antiferromagnetic domain state. In this case, uncompensated spins at domain walls would give rise to an excess magnetization. The Curie-like increase of the magnetization at very low temperatures (below $T$ $\sim$ 25K) both in NCMO and GCMO is attributed to paramagnetic Nd and Gd ions respectively.\cite{dupont}\\
To further elucidate the dynamic nature of the magnetization of NCMO, we have performed FC-relaxation experiments, in which the FC magnetization $M_{FC}$ is recorded versus time ($t$) during $t_m$=10000s at $T_m$=35K after different thermal protocols (\textbf{A}, \textbf{B}, \textbf{C} and \textbf{D}). The obtained $M_{FC}$($T$) curves are plotted in Fig. \ref{fig4}(a), and the corresponding relaxation curves $M_{FC}$($t$) are shown in Fig. \ref{fig4}(b); all curves are labelled on the figure according to the thermal protocol (\textbf{A}-\textbf{D}) employed. \textbf{A}: The sample is rapidly cooled (fast cooling rate) to the measurement temperature $T_m$ $<$ $T_N$ in $H$=20 Oe, and after achieving temperature stability ($\sim$20s), the magnetization is recorded vs. time during $t_m$. \textbf{B}: The sample is rapidly cooled to the lowest temperature, and the FC-relaxation collected after re-heating to $T_m$. The temperature dependence of the magnetization is recorded during the re-heating to $T_m$ and above. \textbf{C}: The sample is rapidly cooled to $T_m$, from where the cooling proceeds with a slower rate and the magnetization is recorded on cooling and re-heating to $T_m$, where the FC relaxation is collected during $t_m$. As in \textbf{B}, the magnetization is further recorded during the re-heating to room temperature. \textbf{D} is similar to \textbf{C}, but using a slower cooling rate when initially cooling to $T_m$. As seen in Fig. \ref{fig4}(b), the relaxation curves obtained for experiments \textbf{A} and \textbf{B} are nearly identical, showing again that the cooling rate at all temperatures below $T_N$ is the key parameter of our effect. As observed earlier in Fig. \ref{fig1}(b), the relaxation diminishes when the effective cooling slows down, from experiments \textbf{A},\textbf{B} to experiment \textbf{D}. The relaxation at $T_m$ reflects the evolution of the domain configuration of the AFM state.\\
Neutron powder diffraction studies on a similar charge and orbital ordered CE-Type AFM manganite\cite{radaelli} indicate magnetic disorder in the Mn$^{3+}$ sublattice, associated to domain boundaries breaking the long range orbital ordering. Recent x-ray scattering results\cite{hill} also indicate a partial orbital ordering of the low temperature phase, leading to an orbital domain state. The here observed cooling rate dependence of $M$ below $T_N$ could thus be related to the intrinsic inhomogeneities of the CE-Type structure, and the nucleation or rearrangement of orbital domains and domain walls to accommodate the large contraction of the structure occurring upon cooling. The cooling rate determines the time allowed to the system to accomodate the structural modifications governed by the temperature. In a similar way, one can also perturb the CE-type state by introducing impurities in the structure, for example by replacing some of the Mn cations by Cr or Ru\cite{raveau}. The orbital ordering is again affected, and FM-like correlations induced.\\
The cooling rate dependence of the FC magnetization depicted in Fig. \ref{fig4}(a) is unusual and deserves some additional comments. In experiment \textbf{B}, the magnetization curve always lies above the curve obtained for the slow cooling case, closely following the curve obtained for a fast cooling. In experiments \textbf{A} and \textbf{C} instead, $M_{FC}/H$ at $T_m$=35K amounts to 0.0185 [SI] immediatly after a rapid cooling from room temperature and the magnetization curve thus surprisingly remains below the FC curve obtained for a slow cooling. During a fast cooling to $T_m$, the system cannot accomodate the excess moments appearing with the domain walls, and its magnetization remains lower than in the slow cooling case. In experiment \textbf{B}, the sample is rapidly cooled down to the lowest temperature, and reaches a high magnetization value. Albeit having a different initial magnetization level, the relaxation curves corresponding to experiments \textbf{A} and \textbf{B} (shown in Fig.\ref{fig4}(b)) appear nearly identical, which indicates that a similar magnetic configuration is probed in both experiments.
  
\section{Conclusion}
The magnetization of a single crystal of the charge ordered manganite Nd$_{0.5}$Ca$_{0.5}$MnO$_3$ is cooling rate dependent below $T_N$. The results reveal the presence of a weak spontaneous moment, related to uncompensated spins in the CE-type AFM structure. The moment and its associated excess magnetization are connected to domain walls separating fully orbital ordered parts of the zig-zag chains of the CE-type structure. The cooling rate is then used to probe the orbital state, and study how it accommodates the large structural transformation occurring upon cooling.\\
The presence of magnetic inhomogeneities in the low temperature CE-type structure of Nd$_{0.5}$Ca$_{0.5}$MnO$_3$ is likely to weaken the AFM state, so that the application of a large magnetic field could induce the observed insulator-metal transition near $T_N$. In the case of Gd$_{0.5}$Ca$_{0.5}$MnO$_3$, no long range antiferomagnetism is established at any temperature, and neither a cooling rate dependence nor an insulator-metal transition is observed. 

\begin{acknowledgments}

Financial support from the Swedish natural Science Research Council (NFR) is acknowledged.

\end{acknowledgments}

%\end{multicols}


\begin{thebibliography}{0}
\bibitem{note}The CO-AFM state of NCMO has been investigated by studying the relaxation of the magnetization and electrical resistance after large magnetic field changes in reference [\onlinecite{kimura}].
\end{thebibliography}
\end{document}